\documentclass[fleqn,10pt]{wlscirep}
\usepackage[utf8]{inputenc}
\usepackage[T1]{fontenc}
\usepackage{comment}

\title{Mixing of a binary passive particle system using smart active particles}

\author[1,2]{Thomas Jacob}
\author[1]{Siddhant Mohapatra}
\author[1]{Rajalingam A}
\author[3]{Sam Mathew}
\author[1,*]{Pallab Sinha Mahapatra}

\affil[1]{Department of Mechanical Engineering, Indian Institute of Technology Madras, 600036 Chennai, India.}
\affil[2]{Department of Mechanical Engineering, Mar Athanasius College of Engineering, 686666 Kothamangalam, India.}
\affil[3]{Fachbereich Physik, Freie Universität Berlin, Arnimallee 14, 14195 Berlin, Germany.}
\affil[*]{email: pallab@iitm.ac.in}


\begin{abstract}

Controlled activity of active entities interacting with a passive environment can generate emergent system-level phenomena, positioning such systems as promising platforms for potential downstream applications in targeted drug delivery, adaptive and reconfigurable materials, microfluidic transport and related fields. The present work aims to realise an optimal mixing of two segregated species of passive particles by introducing a small fraction of active particles ($2\%$ by composition) with adaptive and intelligent behaviour, directed by a trained Artificial Neural Network-based agent. While conventional run-and-tumble particles can induce mixing in the system, the smart active particles demonstrate superior performance, achieving faster and more efficient mixing. Interestingly, an optimal mixing strategy doesn’t involve a uniform dispersion of active particles in the domain, but rather limiting their motion to an eccentrically placed zone of activity, inducing a global rotational motion of the passive particles about the system centre. A transition in the directionality of the passive particles' motion is observed along the radius towards the centre, likening the active particles' motion to an ellipse-shaped void with a defined surface speed. Situated at the intersection of active matter and machine learning, this work highlights the potential of integrating adaptive learning frameworks into traditional active matter models.
\end{abstract}
\begin{document}

\flushbottom
\maketitle

\thispagestyle{empty}

\section*{Introduction}

Active matter encompasses broad classes of intrinsically non-equilibrium systems, forming a ubiquitous part of natural as well as synthetic systems across size scales. These systems consist of units which constantly dissipate energy and, through local interactions, give rise to emergent system-spanning behavioural patterns, otherwise termed collective behaviour. Such system behaviour has been reported in several experimental forays on natural \cite{ballerini2008interaction, cook2020individual, mattingly2022collective} and artificial \cite{duarte2016evolution,mezey2025purely} systems. The 1990s marked the development of the mathematical machinery to understand and interpret the active systems with pioneering works by Vicsek et al. \cite{vicsek1995novel}, and Toner and Tu \cite{toner1995long}. These seminal works led to an increased interest in the numerical modelling of active systems and their application to a wide range of fields encompassing life sciences, physical sciences, safety science, and econometrics. Reviews by Ramaswamy et al. \cite{ramaswamy2010mechanics}, Vicsek and Zaefiris \cite{vicsek2012collective}, Marchetti et al. \cite{marchetti2013hydrodynamics}, and more recently, by Shaebani et al. \cite{shaebani2020computational}, and Gompper et al. \cite{gompper20252025} provide a detailed account of theoretical paradigms and numerical approaches that have shaped the current research in active matter. In the biological world, evolutionary pressures driven by basic functional needs are considered central to the emergence of social behaviour in organisms. Some instances of this social behaviour among macroscale organisms include flocking/swarming for predator confusion \cite{hogan2017confusion, mohapatra2019confined, olson2013predator, mohapatra2025behavioural}, navigation of complex environments \cite{cavagna2013diffusion}, schooling for hydrodynamic efficiency \cite{saadat2021hydrodynamic}, and herding for coordinated escape \cite{cressman2011effects}. In the microscopic world, collective behaviour is often driven by physico-chemical processes or biological cues, some examples of which are the chemotaxis of \textit{Escherichia coli} causing swarming \cite{colin2019chemotactic}, the mechanotaxis of \textit{Pseudomonas aeruginosa} through twitching mobility leading to the formation of rafts \cite{kuhn2021mechanotaxis}, and the chemical gradient-driven active nematic motion of rod-like cells resulting in the formation of lanes \cite{memarian2021active}. Irrespective of the size scales, the collective behaviour observed in active systems stems purely from localised interactions involving proximate entities. In such dynamic systems, natural or artificial, the active entities often interact with passive structures/entities.\cite{wu2000particle, volpe2011microswimmers, angelani2011effective, valeriani2011colloids, schwarz2012phase}. The scope and versatility of the phenomena deriving from such interactions, most notably clustering, homogeneous mixing, phase separation, and active transport, have prompted extensive studies on active-passive mixtures of varying size ratios \cite{dolai2018phase}, activity \cite{mccandlish2012spontaneous, gokhale2022dynamic, dhar2024active}, and particle proportions \cite{hrishikesh2022collective}. Experiments introducing a minute fraction ($\approx 1\%$ by area) of active particles in a dense aggregation of passive colloids (varying between $10\%$ to $90\%$ by area) demonstrated that even a highly limited active component can significantly alter the structure and dynamics of the system \cite{kummel2015formation}. Microscopic parameters such as particle activity and interaction strengths, as well as macroscopic properties such as particle concentration, have been found to affect the emergent behaviour in active and active-passive systems, raising pertinent questions about the parameter space in which such systems operate.

Recent advancements in Machine Learning (ML) have provided a bottom-up approach to understanding active systems. Supervised as well as unsupervised learning techniques have been deployed in various applications involving active matter, such as pattern recognition and classification \cite{mototaketopological, nishida2018robust}, predictive modelling \cite{dulaney2021machine, colen2021machine}, optimal navigation strategies \cite{colabrese2017flow}, and swarm optimisation \cite{loffler2023collective}. Several studies demonstrate the efficacy of reinforcement learning (RL) in discerning the optimal parameter set of active systems driven towards a specific objective. In the purview of control on a particle scale, a widely studied problem is that of optimal navigation of the particle towards a target location under different environmental conditions such as complex flow fields \cite{colabrese2017flow, gunnarson2021learning, liebchen2019optimal, nasiri2022reinforcement}, spatially varying motility landscapes \cite{monderkamp2022active}, physical or potential barriers \cite{schneider2019optimal}, and stochasticity in the surroundings \cite{nasiri2024smart}. Some studies have also focused on controlling the particle motion through selective activation using attraction/repulsion forces \cite{schildknecht2022reinforcement}, optical exposure \cite{falk2021learning, loffler2023collective}, among others. Therefore, a system of randomly interacting particles could be trained to be more efficient in achieving the desired goals by controlling one or more parameters of the system- usually speed and/or direction.

In the literature, Q-learning, a reinforcement learning algorithm, stands out as a prominent tool for training relatively less complex problems such as grid world navigation\cite{antony2023q}, maze solving\cite{tijsma2016comparing}, and path planning \cite{schneider2019optimal,  colabrese2017flow, mirzakhanloo2020active}, often using Q-tables for state-action mapping. However, with problems requiring a larger state action space (find technical details in the Methods section), increased dimensionality, and larger exploration, the learning gets progressively complex and cannot be handled by Q-tables. Artificial Neural Networks (ANN) have to be employed to handle the increased complexity. Deep Q Networks (DQN) is one of the most successful initial implementations using a Deep Neural Network (DNN), with a value-based off-policy algorithm. Since the advent of DQN, several algorithms have emerged to train deep neural networks effectively. These algorithms can be classified as value-based (e.g., DQN, DDQN), policy-based (e.g., REINFORCE), and actor-critic based (e.g., PPO, SAC). Each of these algorithms, while being successful for certain problems, also presents its own set of challenges. Due to the decision-making role played by the agent when integrated with active matter systems (such as propulsion/directional control), the selection of the type of algorithm is contingent upon the characteristics of the action space and the nature of the control problem. Discrete action scenarios are the primary application of value-based algorithms, such as DQN. Policy-based as well as hybrid (combining policy and value) algorithms can generally handle continuous as well as discrete action spaces. It is noteworthy that value-based algorithms have been predominantly applied to single-particle manoeuvring. However, in systems with multiple particles, it is preferable to switch to hybrid algorithms which use an actor-critic framework. 

In the current study, a small number of active particles are introduced into a two-dimensional binary athermal bath, consisting of two initially segregated passive species, inside a circular confinement. The objective of the active particles is to agitate the passive particles and achieve an efficient mixture of the species. The motion of the active particles is controlled by an agent which has been trained through reinforcement learning to maximise an objective function (defined as a mixing index of the passive system). The present work involves passive particles driven by direct impact with active particles, in contrast to extant literature \cite{schildknecht2022reinforcement, falk2021learning, loffler2023collective} which employed various methods to induce self-propulsion in the targeted particles. Additionally, the physical properties of both the passive species are assumed to be similar, for the sake of simplicity; therefore, they are differentiated only by colour. The complexity of the problem arises in the training stage, due to the large number of possible actions in any given system state, especially when the objective function is a macroscopic quantity of the entire system. Considering the highly non-linear nature of mapping the system states to probable actions, the method demands a non-conventional approach involving Artificial Neural Networks (ANN). Using Reinforcement Learning (RL) concepts, in which an agent learns to achieve an objective through repeated experiences, the current work demonstrates that a minute fraction of active particles trained to perform simple discrete actions suffices to mix a binary passive system efficiently. The next section details the numerical methodology, including the equations of motion governing the active and passive particles, and the reinforcement learning framework. It also describes the method of quantifying mixing among passive particles, which is further used to define the objective function of the optimisation problem. The subsequent sections pertain to the mixing performance of active particles following run-and-tumble dynamics, before discussing and comparing it to the mixing efficacy when employing active particles controlled by a trained RL agent.

\section*{Numerical methodology} \label{Method}
\subsection*{Simulation environment} \label{SE}

\begin{figure}[ht]
\centering
\includegraphics[scale=0.7]{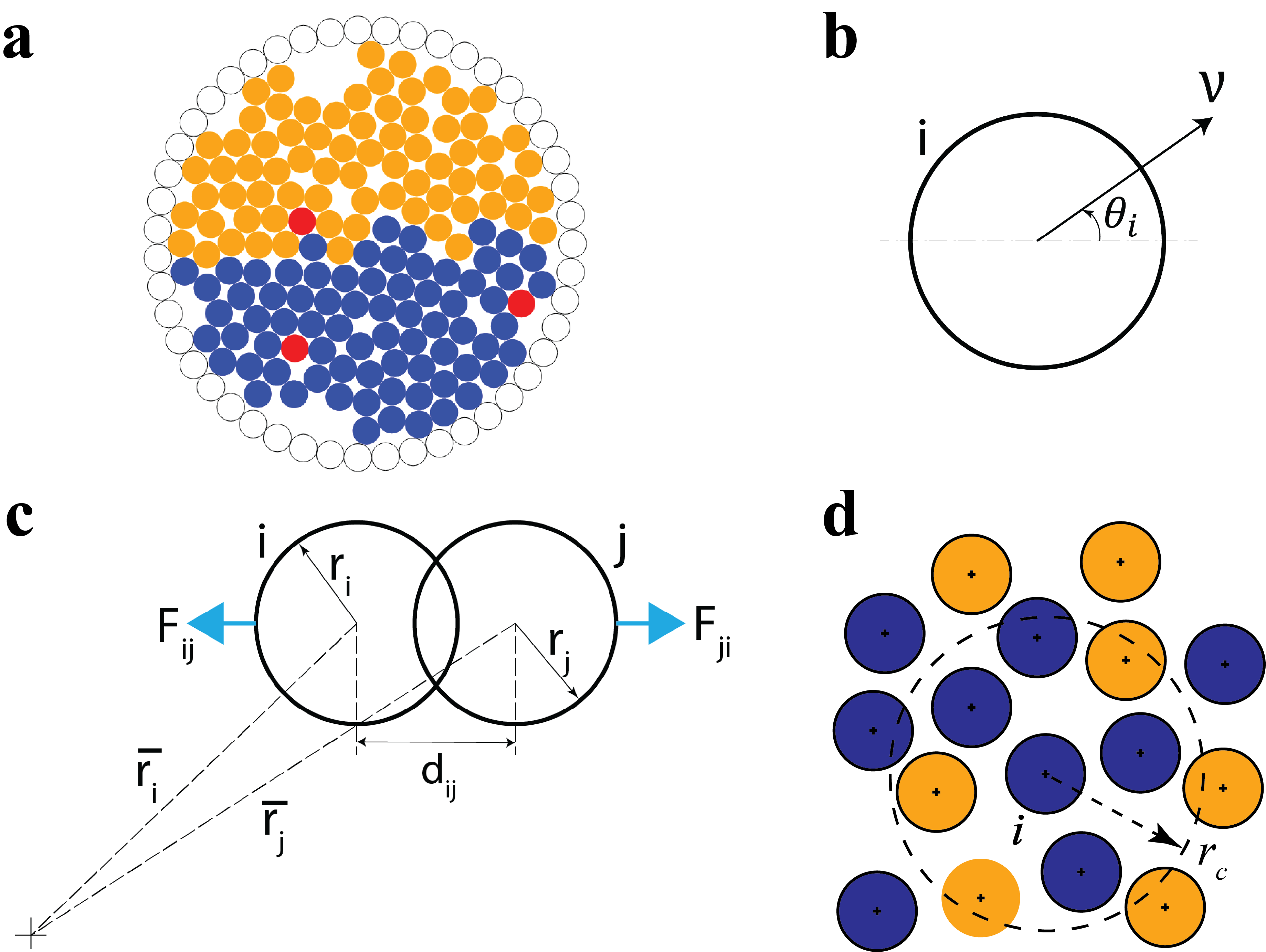}
\caption{Panel (a) showcases the initial distribution of particles in the confined circular domain. The active particles are coloured red, while the two passive particle species are coloured blue and orange. The wall particles are demarcated by their circumference in black. Ratio of radius of domain to that of a passive particle $\rho = 15$, while the packing fraction of passive particles $\phi_{p}=0.65$, and the number of active particles $N_{a}=3$. Panel (b) displays the self-propulsion drive of any active particle $i$ in the direction $\theta_{i}$ at speed $v$, while panel (c) illustrates the inter-particle repulsion drive $F_{ij}$ and $F_{ji}$ acting on particles $i$ and $j$, respectively, on overlap. The strength of this drive scales linearly with the extent of overlap. Panel (d) illustrates the neighbourhood selection scheme for any particle $i$ in the calculation of the mixing index. A metric-based neighbourhood selection is used, when any particle within a distance of $r_{c}$ from the centre of $i$ is considered a neighbour of particle $i$. According to the example in panel (d), particle $i$ has seven neighbours ($n=7$), four of which are of the same type (blue) as $i$, while the rest are of the opposite type (orange; $n_{o}=3$). Hence the number fraction of opposite species for particle $i$ is $\zeta=\frac{n_{o}}{n}=\frac{3}{7}$.}
\label{fig:ini_dist_sp}
\end{figure}

Monodisperse passive disc-shaped particles are uniformly distributed inside a confined two-dimensional circular domain as represented in Fig. \ref{fig:ini_dist_sp}(a). All particles have the same radius $r$, and the ratio of the domain radius $R$ to the particle radius is defined as the radius ratio $\rho$. The wall of the bounded system consists of one layer of circular discs of the same size as the interior particles. The packing in the system is denoted by an area fraction $\phi=Nr^2/R^2$, where $N$ is the number of interior particles. It is to be noted that the packing fraction takes into account the passive particles only, while the active particles are represented in numbers, $N_{a}$.

The governing equations of motion of the active and the passive particles are delineated in Eqs. \ref{active_dyn} and \ref{passive_dyn}, adopted from the model used by Henkes et al. \cite{henkes2011active}. The current model assumes non-inertial and athermal particles and is valid for particles moving slowly or in highly viscous environments (such that inertial effects are negligible in comparison to viscous damping). The active particles are acted upon by two drives: the self-propulsion drive and the inter-particle repulsion drive (see Eq. \ref{active_dyn}).

\begin{equation} \label{active_dyn}
\mathbf{\dot{x}_i}=v\hat{\mathbf{n}}_i + \mu \sum_{j} \mathbf{F}_{ij}
\end{equation}
\noindent Here, $\mathbf{x_{i}}$ is the position of active particle $i$ with respect to the origin, $v$ is the self-propulsion speed in the direction $\hat{\mathbf{n}}_i$, $F_{ij}=k(r_{i}+r_{j}-d_{ij}) \quad\textbf{if} \quad r_i + r_j > d_{ij}$ is the repulsive force exerted on particle $i$ due to overlap with any particle $j$, $\mu$ is the translational mobility, $k$ is the coefficient of the repulsive force, $r_{i}$ and $r_{j}$ are the radii of the particles $i$ and $j$, respectively. $d_{ij}=\|\mathbf{x_i}-\mathbf{x_j}\|$ is the distance between the centres of particles $i$ and $j$. The direction of motion of the active particle $i$ is controlled by the term $\mathbf{\hat{n}_i} = \begin{pmatrix} \cos \theta_{i} \\ \sin \theta_{i} \end{pmatrix}$, where $\theta_{i}$ is the angle subtended by the desired direction of propulsion of the particle $i$ with the x-axis (see Fig. \ref{fig:ini_dist_sp}(b)). $\theta_{i}$ can change continuously over the range $[0,2\pi)$, or assume discrete angles based on prevalent movement criteria. By modulating the direction of propulsion $\theta_{i}$, different types of motion can be observed in the active particles, such as run-and-tumble (RT) \cite{lee2019computational, junot2022run}, run-reverse \cite{grossmann2016diffusion, guseva2025advantages}, and directed migration \cite{schakenraad2020topotaxis, bhattacharjee2022chemotactic}.

\begin{equation}
\label{passive_dyn}
\mathbf{\dot{x}_p}=\mu \sum_{j} \mathbf{F}_{pj}
\end{equation}

The equation for passive particles (Eq. \ref{passive_dyn}) differs from their active counterpart due to their inability to self-propel. Therefore, the passive particles are subjected only to the inter-particle repulsion drive. In the current work, all particles are assumed to be athermal (i.e., possessing no thermal diffusivity).

To observe mixing, the passive particles are initially segregated along the diameter of the circular domain, while the active particles are uniformly distributed (see Fig. \ref{fig:ini_dist_sp}(a)). In the present study, unless otherwise specified, three active particles ($N_{a}=3$) are used to agitate a relatively dense binary passive aggregation ($\phi_{P}=0.65$). All particles are assumed to be of unit radius, and any length dimensions presented are scaled against it. To keep the system concise and manageable, the radius ratio $\rho$ is fixed at $15$, $\mu k=10$, and $v=1$. Simulations are carried out using the first-order Euler explicit time integration scheme with a time step of $\Delta\tau=0.01$. The results presented in the next section pertain to $\tau=5\times10^{6}$ time steps or longer (to observe typical long-time behaviour). The upcoming subsection explains the dynamics of run-and-tumble particles (RTPs) and their efficacy in mixing the two passive species.

\subsection*{Run-and-tumble particles}

Run-and-tumble (RT) is one of the prevalent mechanisms of bacterial locomotion, often observed in species such as \textit{Escherichia coli}, \textit{Bacillus subtilis}, and \textit{Salmonella enterica}. RT motion is characterised by ballistic ``runs", interspersed by sudden directional changes (tumbles). Empirical evidence of locomotion of these microbes postulates an exponential distribution of the run duration, with tumble angles uniformly distributed about a range. However, for artificial systems, the tumbling range can be $[0,2\pi)$, which induces complete randomness. The active particles, having been modelled as slow-moving robots, can be thought to behave as RT particles with a certain mean run duration and instantaneous tumbles at any angle in the range $[0,2\pi)$. This enhances the exploration probability over the entire domain due to the synergy between persistent runs and random tumbles. The dynamics of these particles are governed by Eqs.\ref{active_dyn} and \ref{passive_dyn}, and the run duration is governed by Eq.\ref{rt_exp_dist}, where $\tau_{r}$ is the sampled run duration and $\tau_{m}$ is the mean run duration.
\begin{equation} \label{rt_exp_dist}
\mathbf{\tau_{r}}=\frac{1}{\tau_{m}}e^{\frac{-t}{\tau_{m}}}
\end{equation}

\noindent In the current work, run-and-tumble particles are employed to behave as a randomised mixer of the passive species, with the mean run duration serving as a primary control parameter. Such an analysis provides a base case for defining programmed mixing functionality with the help of Reinforcement Learning (RL) later on.

\subsection*{Mixing index ($\chi$)} \label{RF}

As discussed previously, the active particles serve to agitate the passive species in the system, promoting their mixing. Therefore, the mixing in the system has to be properly quantified. From the literature, there exist different methods to quantify the mixing of a binary particle system \cite{doucet2008measure, bhalode2020review}, and the choice of the method has to be simple and effective. Due to the inherent large fluctuations observed in quantifying mixing while dealing with grid-based methods, it was ruled out. Mixing can also be computed based on Principal Component Analysis (PCA) \cite{doucet2008measure}; however, it is computationally expensive. In granular mixing, several methods have been developed to assess the extent of mixing in a binary particle system. In the current work, a relatively straightforward and computationally efficient method has been used to quantify the mixing index $\chi$, as elucidated in Eq. \ref{MI}.

\begin{equation}
\label{MI}
\chi=\frac{1}{N_{p}}\sum_i^{N_p} (2\zeta_i)
\end{equation}
\noindent Here, $\zeta_i=\frac{n_{o}}{n}$, where $n_{o}$ is the number of passive particles of the opposite species and $n$ is the total particles surrounding the passive particle $i$, within a fixed radius $r_{c}$ from the centre of $i$ (see Figure \ref{active_dyn}(d)). Only passive particle species are considered to compute the mixing index. For a dense system, the value of $\chi$ varies from a value greater than $0$ for an unmixed system (in the initial positional configuration, the particles at the interface of the two species have non-zero $\zeta$) to a value close to $1$ in a well-mixed system. Fluctuations are, however, possible even in well-mixed systems in the case of sparse packing (lower value of $\phi_{p}$). In the current study, the passive particles present a relatively dense system ($\phi_{p}=0.65$) and Eq. \ref{MI} is applied in a neighbourhood definition of radius $r_{c}=4r$. For an ideally mixed system, there will be an equal number of opposite species surrounding every passive particle, bringing the ratio $\zeta=0.5$ and averaging it across all passive particles results in $\chi=1$.

\subsection*{Reinforcement learning framework}

\begin{figure}[ht]
\centering
\includegraphics[scale=0.65]{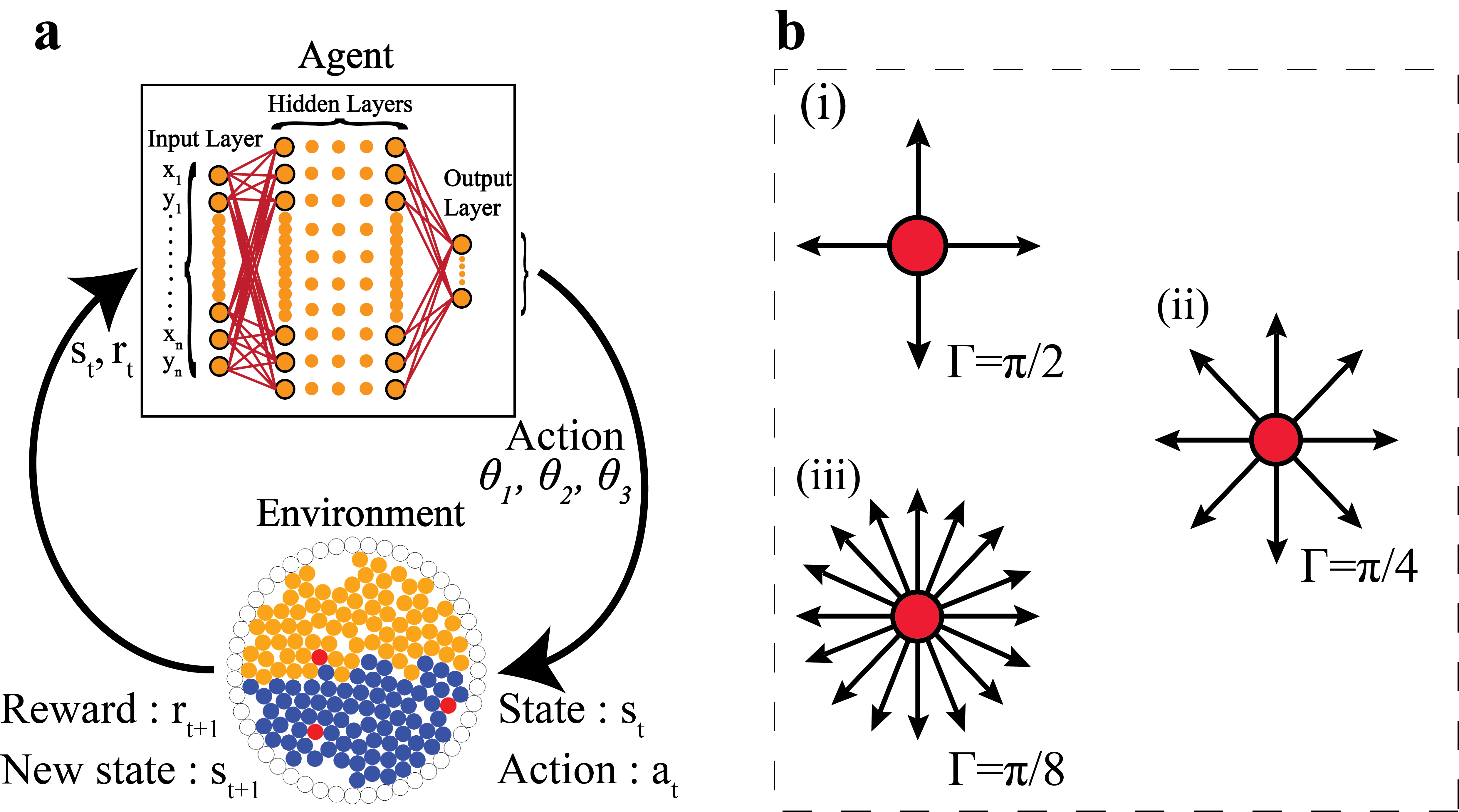}
\caption{Panel (a) illustrates the interaction between the reinforcement learning agent and the environment. Here, the agent refers to an Artificial Neural Network (ANN). Action $a_{t}$ transforms the state of the environment from $s_{t}$ to $s_{t+1}$ along with a feedback in the form of a reward $r_{t+1}$. Panel (b) demonstrates the different steps $\Gamma$ by which the RL agent can modulate the direction of motion for the active particles. The orientation of the active particles can be modulated in steps of (i) $\Gamma=\pi/2$ ($4$ possible directions of motion), (ii) $\Gamma=\pi/4$ ($8$ possible directions of motion), and (iii) $\Gamma=\pi/8$ ($16$ possible directions of motion).}
\label{fig:agent_env}
\end{figure}

The Reinforcement Learning (RL) framework is employed to find an optimal mixing strategy to efficiently mix the passive system by guiding active particles. Here, optimality refers to the shortest path (lowest simulation time) to a high value of mixing index $\chi$. As discussed earlier, active particles can be controlled by adjusting their run duration $\delta$ and direction of motion $\theta_{i}$ in steps of $\Gamma$. Figure \ref{fig:agent_env}(a) illustrates the RL setup consisting of two key components: the agent and the environment, interacting through three quantities: the state (observation), the action and the reward. The agent is the decision maker, and the environment is the system whose state the agent attempts to modulate. This modulation is made possible by communicating action variables $a_{t}$ to the environment, based on the current state of the environment at any time $t$ (also called the observation $s_{t}$). Due to the implementation of the action, the environment undergoes a transition to a new state $s_{t+1}$. This transition to the new state concurrently results in a reward value $r_{t+1}$, which is usually based on $s_{t}$ and $s_{t+1}$. The reward value is quintessentially a quantification of the effect of the action $a_t$ on the state $s_t$. The result is the formation of a tuple $\left(s_{t}, a_{t}, r_{t+1}, s_{t+1} \right)$. In the current study, the mixing index $\chi$ (refer to the previous subsection) constitutes the reward function. The coordinates for the active as well as the passive particles, form the observation space in the current RL framework - the input is in the form of a flattened 1D array \{$x_{1},y_{1},x_{2},y_{2},\cdots x_{N_p+N_a},y_{N_p+N_a}$\}. Cyclical interaction between the agent and environment gives rise to a collection of state-action-reward tuples. A large set of possible actions at any state corresponds to a large number of state-action combinations for the active particles, increasing the difficulty of training the agent to guide the environment to an optimal state.

In the current RL training module, the orientations of the active particles ($\theta_{1},\theta_{2},\theta_{3}$) are set to be the action variables communicated by the agent (see Fig. \ref{fig:agent_env}(a)). Although the directional orientations can ideally be set as continuous variables in the range $[0, 2\pi)$, a discrete action space is chosen for the ease of implementation (significantly faster training due to a smaller action space and negligible difference in the final state of the environment). As a result, when a certain action input is provided to an active particle, the particle continues to move in that direction until it receives another action input from the RL agent. In the purview of the current work, an RL agent-controlled active particle is termed a Smart Active Particle (SAP), and the terms ``environment" and ``system" are used interchangeably. To allow adequate time for the SAPs to interact and mix the passive particles, the agent transmits action variables to the SAPs every $\delta$ time steps (also known as the run duration for the SAPs, inspired by the RT dynamics). After every run duration, each SAP tumbles instantaneously to a new orientation. The tumbled (new) orientations of the active particles are the sum of the previous orientations and the action variables transmitted from the RL agent. As the action space is discrete, the tumbling can occur in steps of $\Gamma$, bringing the number of possible actions for an SAP at any state to $2\pi/\Gamma$. Therefore, the number of combinations of the possible actions for $N_a$ SAPs in any state amounts to $(2\pi/\Gamma)^{N_a}$. The lower the value of $\Gamma$ or the higher the number of SAPs, the larger the action space, with $N_{a}$ being the greater influence of the two. Figure \ref{fig:agent_env}(b) illustrates the three values of $\Gamma=\{\pi/2, \pi/4, \pi/8\}$ tested in the current work, corresponding to $4$, $8$, and $16$ possible action directions for each SAP, respectively. Additionally, taking into account a fairly dense passive aggregation, the observation space turns out to be large enough to warrant the use of an ANN with multiple hidden layers for representing the RL agent (with a shared network for both policy and value functions). The parameters of the ANN are randomly initialised and are updated throughout the course of the training. A detailed description of the RL implementation is discussed in Sec. SI-1 and SI-2 of Supplementary Information, and visualised in Fig. S1 of Supplementary Information.

A MultiLayer Perceptron (MLP) policy with a ReLU activation function is selected, from a wide variety of ANNs, to represent the agent in the RL implementation. ReLU provides nonlinearity to the neural network, enabling it to learn complex mappings between action probabilities and state inputs. Proximal Policy Optimisation (PPO) is used to optimise the parameters of the MLP policy due to its stability in updating network parameters from a clipped surrogate objective function, its capability to manage discrete action spaces, the sample efficiency \cite{schulman2017proximal}, and the effectiveness in addressing physical problems related to active matter and optimal navigation. The policy update is executed using the PPO algorithm with the primary aim of maximising the cumulative reward (or minimising a loss function). The approach aggregates a sequence of tuples prior to policy update throughout the experience collection process. The whole reinforcement learning framework is constructed within the OpenAI Gym interface with the \textit{stable-baselines3} package in Python. Among the several hyperparameters in PPO, the learning rate is one of the most crucial in influencing the training efficacy in terms of convergence and speed. It controls the extent to which the policy's weights and parameters are adjusted in response to the computed policy gradient. Preliminary simulations indicate a lower than default learning rate of $10^{-6}$ to avert unstable parameter updates in our policy (see Sec. SI-3, Fig. S2, and Tables S1 and S2 of Supplementary Information for the preliminary simulation details and the selection of the hyperparameters). Following a detailed investigation of the impact of $\Gamma$ and $\delta$ in conjunction with the chosen learning rate, and taking into account several hidden layer configurations for the MLP policy, a neural network with hidden layer sizes of $(512,256,64)$ has been selected to represent the agent (refer to Sec. SI-3 and Figs. S3 through S5 of Supplementary Information for details).

\section*{Results} \label{Results}
Prior to examining the dynamics of SAPs and their efficacy in mixing the segregated passive system, it is pertinent to first consider a baseline case without learning, where the mixing arises exclusively by stochastic driving of the active particles. Run-and-tumble particles serve as a useful model for examining self-propelled motion, particularly in the context of translational applications employing microrobots. Therefore, the RT dynamics serve as a paradigm for the development of SAPs, which can be programmed to perform certain tasks in designated environments. The upcoming subsection analyses the influence of active particles following RT dynamics on the passive species in a confined circular domain, which is subsequently contrasted against SAPs.

\subsection*{Mixing by run-and-tumble particles}

\begin{figure}
\centering
\includegraphics[scale=0.50]{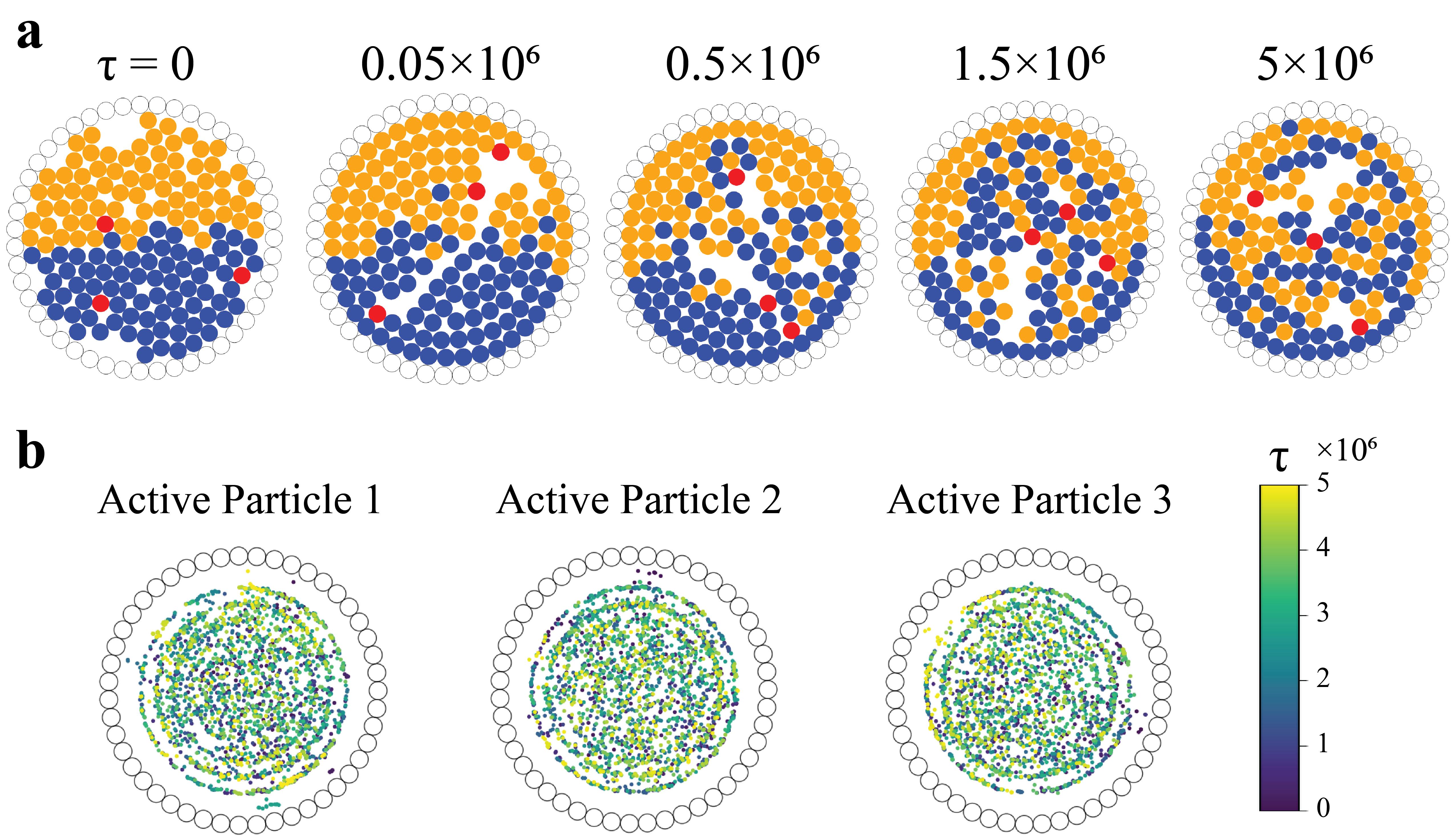}
\caption{Panel (a) displays snapshots of the system at different time instances $\tau$. The active particles are coloured red, while the two passive species are coloured orange and blue, respectively. Panel (b) showcases the locations (coloured by time stamp) of the tumbling events of the active particles following run-and-tumble (RT) dynamics. (Note: Both the panels involve depiction of a representative simulation, and qualitatively similar features are observed across multiple realisations.)}
\label{fig:RT_Mixing}
\end{figure}

\begin{figure}
\centering
\includegraphics[scale=0.80]{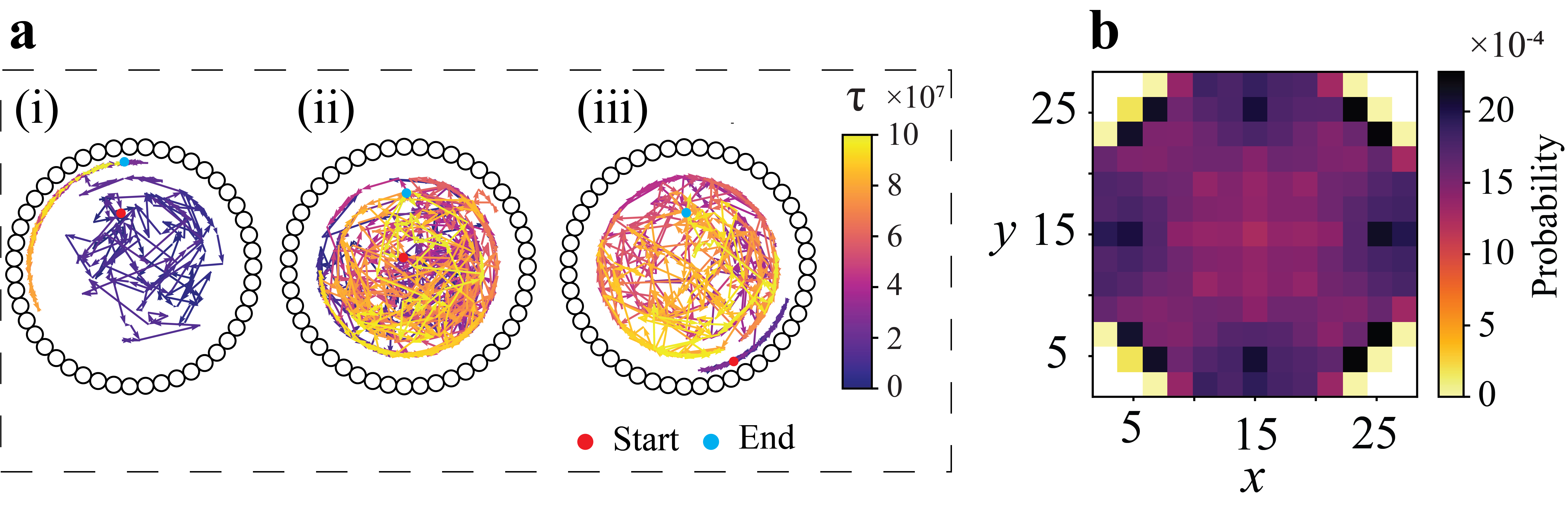}
\caption{The trajectories of three representative passive particles are displayed to showcase their long-term behaviour, on interaction with run-and-tumble active particles (time evolution is represented through a colour gradient). The passive particles are chosen based on their initial positions in the domain: (a-i) located somewhat off-centre, (a-ii) centrally located, and (a-iii) located next to the wall. The starting point of these passive particles is marked with red disks, while the final position (i.e., $\tau=10^8$ time steps) is marked with cyan disks. The arrows represent the direction of motion of the particles at any point. Panel (b) presents the spatial probability distribution of all positions occupied by the passive particles over a long time window ($0$ to $\approx 10^8$ time steps).}
\label{fig:RT_traj}
\end{figure}

Inspired by several microscopic organisms, run-and-tumble (RT) dynamics for active particles serve as a benchmark for highlighting the mixing performance of active particles. Figure \ref{fig:RT_Mixing}(a) provides a visual depiction of a typical mixing scenario where the active particles follow RT dynamics interacting with the stratified binary passive system over a period of time ($5\times10^{6}$ time steps). The presented case involves a mean run duration of $\tau_m=2\times10^{3}$ time steps and the angle of tumble uniformly distributed in $[0, 2\pi)$. A gradual mixing can be observed in the series of time progression snapshots of the system, arising from interactions among the active particles and the two passive species. Figure \ref{fig:RT_Mixing}(b) showcases the spatial mapping of the locations of tumble events for the RT particles, coloured by time stamp. It is clear that all the active particles explore the entire domain, sans the region immediately adjacent to the wall. Simulating random tumbles, as is the case with RT particles, the active particles barely perturb the passive particles along the wall, which explains the absence of tumbling events adjacent to the wall. Although Fig. \ref{fig:RT_Mixing} demonstrates the features of a representative case, qualitatively similar behaviour is observed across multiple realisations and different simulation parameters. To understand the effect of the RT particles on the behaviour of the passive particles, the trajectories of three passive particles from different locations (centre, off-centre, and next to the wall) are showcased in Figs. \ref{fig:RT_traj}(a(i--iii)) over a sufficiently long simulation time ($\approx 10^{8}$ time steps). Following the time signature in the form of the colour gradient, it is observed that the passive particles are driven through the domain in a random fashion. To bolster these observations, Fig. \ref{fig:RT_traj}(b) illustrates the probability of finding any passive particle at any location in the domain, considering all passive particles' positions over a prolonged trajectory ($\tau=10^{8}$ time steps). The highest probability is observed at the domain boundary, arising mainly from increased residence time for the particles near the wall. The probability data also reaffirms an overall stochastic motion for the passive particles in the majority of the domain (similar probability values pointing towards a uniform distribution). The displacement of the passive particles is more pronounced away from the boundaries; hence, the ones travelling along the confinement exhibit minimal positional shifts. An extensive quantitative measurement of the mixing performance of the RT particles has been reported for a range of mean run durations in the upcoming subsection.

\begin{figure}
\centering
\includegraphics[scale = 0.65]{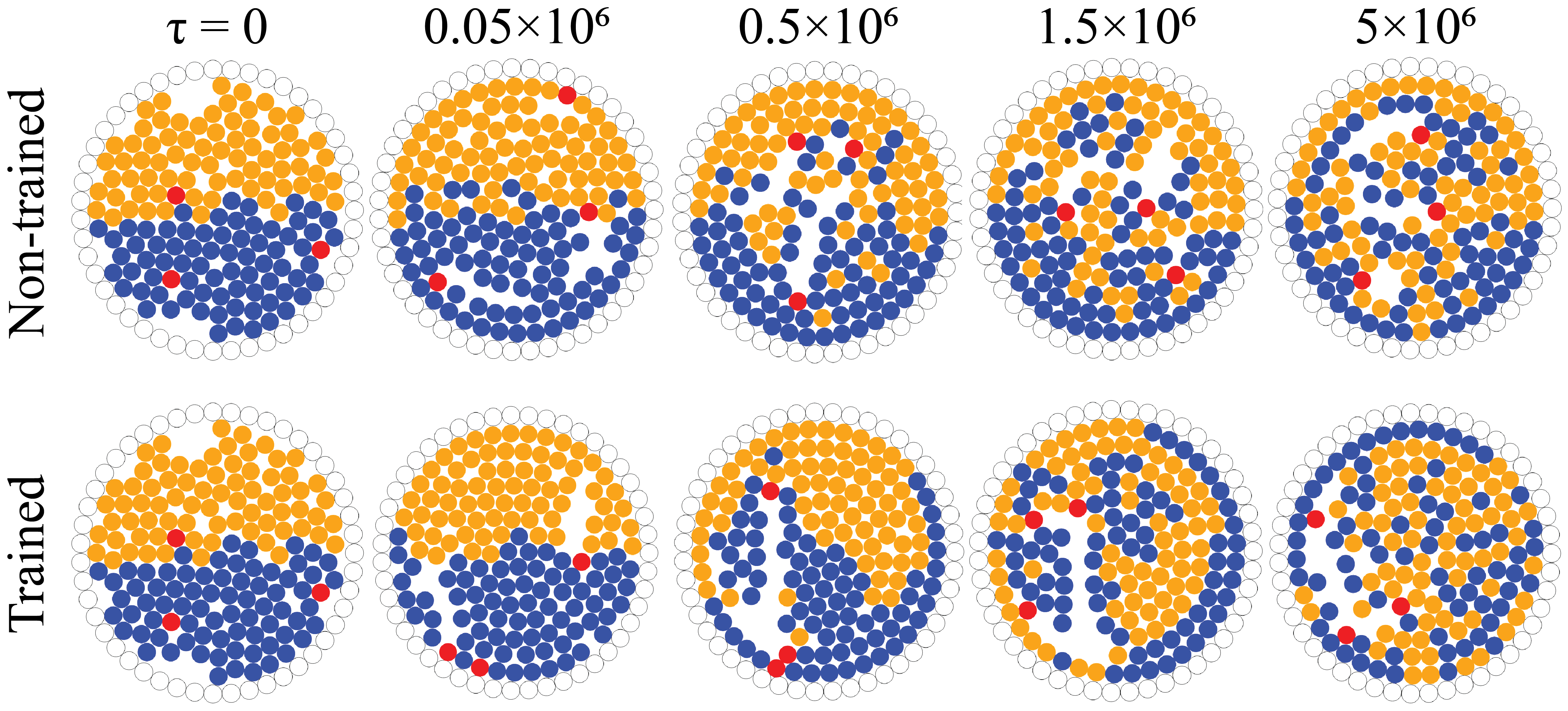}
\caption{Time progression of the mixing of the two passive species enabled by smart active particles (SAPs) controlled by a non-trained agent (top panel) and a trained agent (bottom panel) is presented through snapshots of the domain. The SAPs (coloured red) can only move in the four cardinal directions ($\Gamma=\pi/2$), and are set to tumble instantaneously every $\delta=2\times10^{3}$ time steps. (Note: For the bottom panel, the RL agent has been trained for $10^{6}$ agent-environment interactions. Both the panels are representative of fifty test episodes.)}
\label{fig:L_vs_UL}
\end{figure}

\subsection*{Mixing using trained active particles} \label{TIDF}
Although conventional RT based simulations can achieve mixing in the segregated passive system, there is a scope for improvement, especially with the stochastic inputs to directional changes. Therefore, an RL framework explained in the Methods section is employed to train an agent (ANN) to make informed orientational decisions for the SAPs' movements, thereby promoting mixing.

The RL training architecture (refer to the Methods section and Sec. SI-1 and SI-2 of Supplementary Information) requires two primary inputs for the smart active particles (SAPs): the tumble step $\Gamma$, and the run duration $\delta$. In the current section, the training of the agent is carried out with $\Gamma=\pi/2$ and $\delta=2\times10^{3}$ time steps. The reasoning behind the selection of these specific values has been described in Sec. SI-3, and Figs. S3 through S5 of Supplementary Information. If a SAP is controlled by a trained agent, it is termed a trained SAP (TSAP). If controlled by a non-trained agent, the SAP is termed a non-trained SAP (NTSAP). Figure \ref{fig:L_vs_UL} qualitatively compares the progression in mixing among the passive particles on interaction with the NTSAPs (top panel) and the TSAPs (bottom panel). In the case of the NTSAPs, the mixing is random and doesn't follow any pattern. However, in the case of TSAPs, the snapshots demonstrate the incidence of a global clockwise swirl among the passive particles. Another point to note is the enhanced positional shift of the passive particles near the wall in the presence of TSAPs, hinting at improved mixing capabilities of TSAPs compared to those of NTSAPs. Irrespective of training, the mixing results in the formation of a void in the passive system (see $\tau=5\times 10^6$ of Fig. \ref{fig:L_vs_UL}). A clear distinction, however, lies in the location of this void. In the case of mixing using NTSAPs, the void is usually located in a more central location, compared to a peripheral location in the case of TSAPs.

\begin{figure}[ht]
\centering
\includegraphics[scale=0.75]{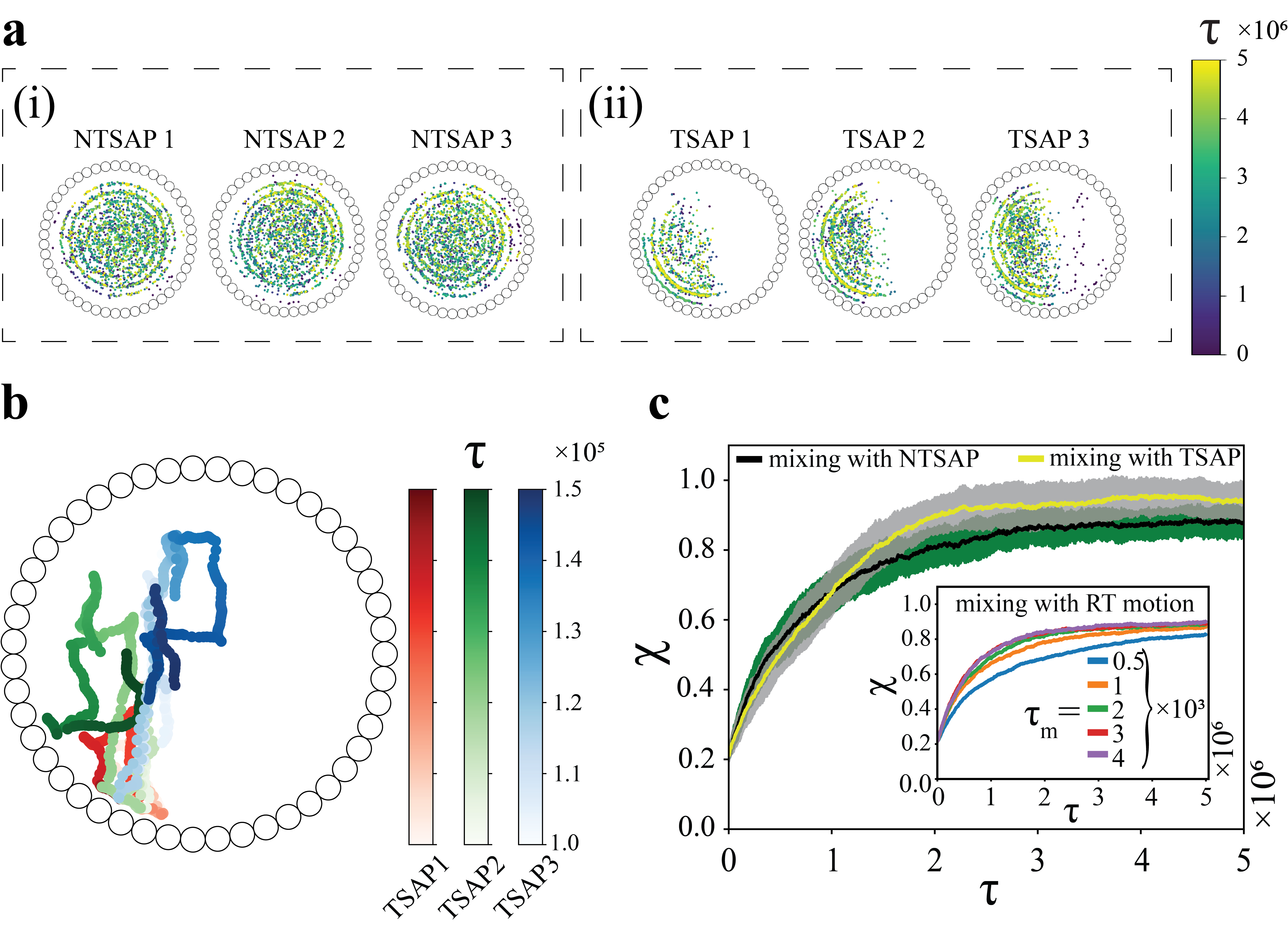}
\caption{Panel (a) illustrates the locations of the tumbling events of the SAPs in the case of (i) a non-trained SAP (NTSAP), and (ii) a trained SAP (TSAP) for a representative test episode. Panel (b) demonstrates the trajectories of three TSAPs for a period of $5\times10^{4}$ time steps, to further highlight the motion of the TSAPs being constrained to only a section of the domain, unlike the NTSAPs. Panel (c) delineates the temporal evolution of the mixing index $\chi$ averaged over fifty test episodes, comparing mixing induced by NTSAPs and TSAPs. The inset in panel (c) represents the change in $\chi$ with time $\tau$ for RT particles with different mean run durations $\tau_{m}$ (refer to Eq. \ref{rt_exp_dist}). All the panels are depicted for parameters $\Gamma=\pi/2$ and $\delta=2\times10^{3}$. (Note: NTSAP refers to a SAP receiving inputs from a non-trained RL agent, while TSAP is a SAP controlled by a trained RL agent.)}
\label{fig:op_vs_nop}
\end{figure}

\begin{figure}[ht!]
\centering
\includegraphics[scale=0.6]{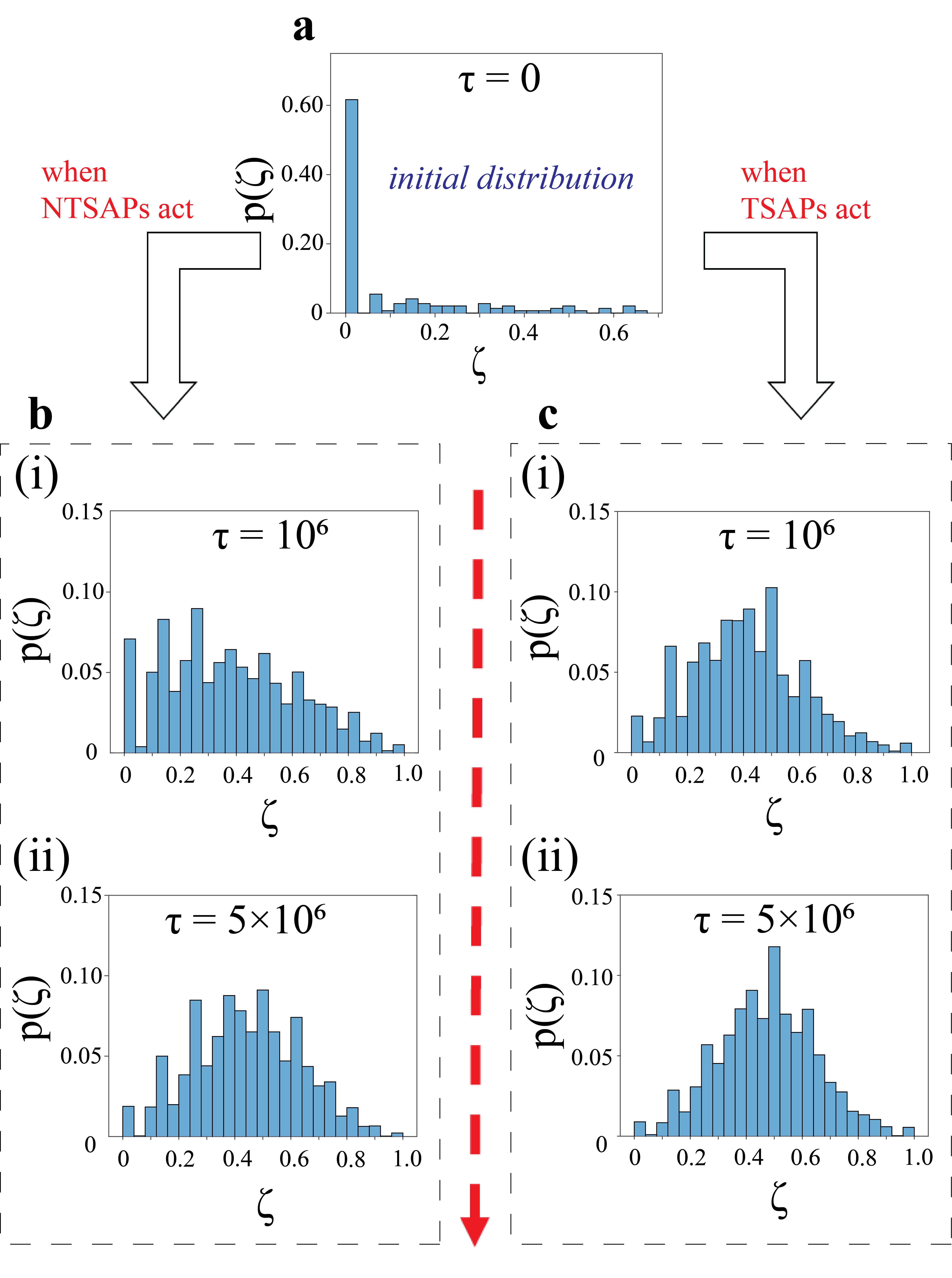}
\caption{The temporal evolution of the probability distribution of $\zeta=\frac{n_{o}}{n}$ (i.e., the ratio of the passive neighbours of dissimilar type/species $n_o$ to the total of passive neighbours $n$ surrounding any passive particle within the radius $r_c$) with (a) initial probability distribution at $\tau=0$ is compared for (b) a system involving NTSAPs, and (c) a system involving TSAPs. The time stamps during evolution are at $\tau=0$, $\tau=10^{6}$, and $\tau=5\times10^{6}$, respectively. The systems using TSAPs for mixing perform better than their counterpart, as shown by the higher probability of finding more dissimilar neighbours at $\tau=10^6$ and $\tau=5\times10^6$, and by the sharper nature of the curve for the bottom panel at $\tau = 5 \times 10^6$.}
\label{fig:PDF_OPP}
\end{figure}

To glean further insights into the mixing phenomena, the locations of the tumbling events of the SAPs have been mapped as shown in Fig. \ref{fig:op_vs_nop}(a). It is evident that a non-trained RL agent prescribes actions which tend to move the active particles randomly across the domain. On the contrary, TSAPs move strategically, focusing on a part of the domain to enhance mixing. The trajectories for the TSAPs are presented over a short time window ($5\times 10^4$ time steps) in Fig.\ref{fig:op_vs_nop}(b), which supports the hypothesis about the constrained motion of the SAPs, characterised by sharp turns ($\Gamma=\pi/2$). The perturbations on the path arise due to collisions with the passive particles and other TSAPs. However, once they reach the periphery, they follow the wall for a duration due to the boundedness and pertaining values of $\Gamma$ and $\delta$. A quantitative measure of the mixing performance is computed with the help of a mixing index defined a priori in the Methods section. Figure \ref{fig:op_vs_nop}(c) compares the temporal variation in the mixing index $\chi$ (ensemble average over fifty test episodes), when mixing is carried out using NTSAPs and TSAPs for a tumble step $\Gamma=\pi/2$ and a run duration $\delta=2\times10^{3}$. In both cases, the mixing index increases with time till it reaches a steady state value. However, a trained agent is seen to be clearly superior in mixing the binary passive system. The steady-state mixing index for the trained RL agent attains a value around $\chi=0.96$, whereas that for the non-trained case stands at $\chi=0.90$. The inset of Fig. \ref{fig:op_vs_nop}(c) illustrates the temporal variation in $\chi$ for mixing using RT particles with different mean run durations $\tau_{m}$, each curve averaged over a hundred realisations. The inset clearly depicts a saturation value close to $\chi=0.90$ for the RT particles with mean run durations beyond $\tau_m=2\times10^{3}$, thereby defining an upper limit to the ability of the RT particles to induce mixing in the binary passive system. The use of a trained RL agent can augment the mixing of the passive system beyond these RTP-based limits, with a substantially simplified approach (restricting the motion of the SAPs to discrete steps in the four cardinal directions).

\subsection*{Probability of finding dissimilar neighbours}
The previous sections have been concerned with the ``microscopic" behaviour of the passive particles; however, it is equally important to understand the ''macroscopic" implications of the actions undertaken by the TSAPs. A well-mixed system, in the context of this passive system, can be postulated to have an equal number of particles of similar and opposite species/type surrounding each passive particle (within the radius $r_{c}$). To test this hypothesis, the probability distribution of $\zeta$, represented as $p(\zeta)$, is computed where the SAPs are not taken into account. In a perfectly mixed dense system with a large number of particles, the histogram should peak at $\zeta=0.5$ with $p(\zeta)=1$, and $p(\zeta) \quad\forall \quad\zeta \neq 0.5$ assuming a value of $0$. 

Figure \ref{fig:PDF_OPP} compares the temporal evolution in the probability density function of $\zeta$, starting from the same initial distribution (Fig. \ref{fig:PDF_OPP}(a)), when using NTSAPs (Fig. \ref{fig:PDF_OPP}(b)) and TSAPs (Fig. \ref{fig:PDF_OPP}(c)) for mixing the passive system. Due to the initially stratified state of the system, all passive particles apart from those at the interface of the two passive species have neighbours of a similar type. Therefore, $p(\zeta)$ has a peak ($\approx 0.6$) at $\zeta = 0$ at $\tau=0$. As time progresses, the SAPs start agitating and mixing the system, causing the peak of $p(\zeta)$ to shift towards higher $\zeta$ values. However, a clear distinction can be made between the performance of the NTSAPs and the TSAPs, at $\tau=10^6$ time steps. Figure \ref{fig:PDF_OPP}(c), illustrating mixing effected by TSAPs, shows a higher probability of finding more dissimilar neighbours around each passive particle, compared to a system where mixing is actuated by NTSAPs. At $\tau=5\times10^6$ time steps, when the mixing index $\chi$ has saturated (see Fig. \ref{fig:op_vs_nop}(c)), the system involving TSAPs outperforms its counterpart, as evidenced by a higher probability of finding more dissimilar particles (the peaks in Figs. \ref{fig:PDF_OPP}(b-ii) and \ref{fig:PDF_OPP}(c-ii) occur at $p(\zeta=0.50) = 0.09$ and $p(\zeta=0.50) = 0.12$, respectively). The distribution of $\zeta$ in the system with the TSAPs closely resembles a Gaussian distribution with a mean at $\zeta=0.47$. Furthermore, on fitting the observed histograms at $\tau=5\times 10^6$ to a Gaussian distribution, the kurtosis points at platykurtic distributions (negative excess kurtosis). A significantly higher negative excess kurtosis is observed in the case of the system with NTSAPs (excess kurtosis of $-0.46$) compared to that of TSAPs (excess kurtosis of $-0.05$), indicating a flatter distribution in the former.

\subsection*{Trajectories for the passive particles with TSAPs}

\begin{figure}[ht!]
\centering
\includegraphics[scale=0.7]{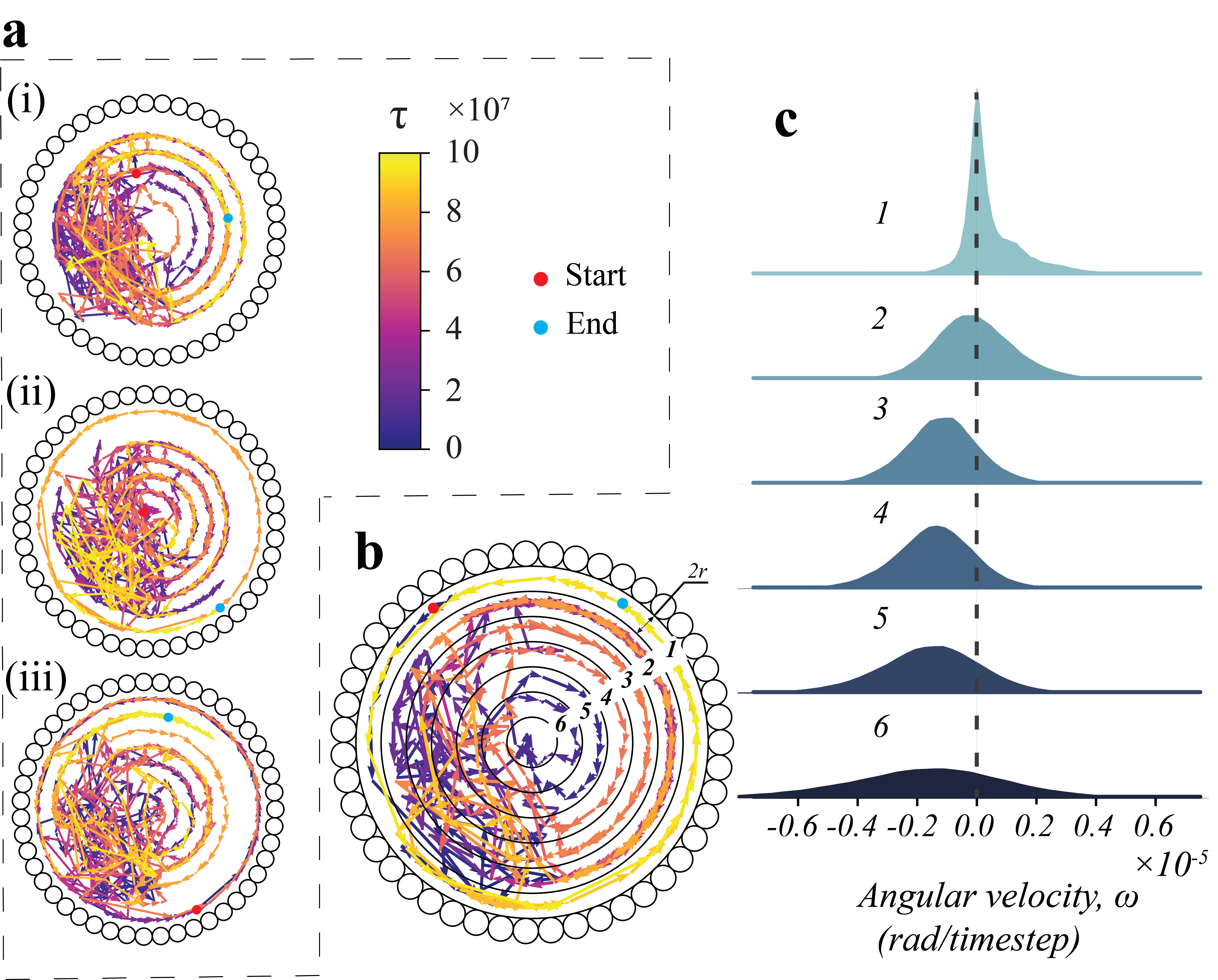}
\caption{Panel (a) illustrates the motion trajectories for three representative passive particles selected based on their initial positions: (a-i) off-centre, (a-ii) centre, and (a-iii) adjacent to the wall. The colour bar indicates the time until a maximum of $\tau=10^{8}$ time steps. The red and the blue disks represent the initial and final positions of the passive particle in each panel. (Note: Each trajectory is expressed from the positions of the passive particle in steps of $10^{5}$ time steps.) Panel (b) features the trajectory (for $10^{8}$ time steps) for a passive particle originating from the periphery (red disk). The entire domain is divided into six concentric regions (sans the central area), numbered $1-6$ starting from the outermost band, with each concentric band of radial width $2r$. Panel (c) showcases the distribution of average angular velocity $\omega$ of the passive particles in each of the bands presented in panel (b) in a system mixed using TSAPs. Each of the simulations is carried out till $\tau=10^{8}$ time steps, and the distribution takes into account the data over fifty test episodes, considering all the passive particles. The black dashed line represents zero angular velocity ($\omega=0$).}
\label{fig:part_pair_evol}
\end{figure}

Visualisation and critical analysis of the dynamics of the passive particles is crucial to understanding the mixing performance of SAPs. Figure \ref{fig:L_vs_UL} provides some insights into the motion of the passive particles as a whole. Visual inspections of the passive system displayed a clockwise swirl about the domain centre in all passive particles except those near the wall (which undergo motion in the anti-clockwise direction). To highlight the aforementioned rotational motion, three passive particles were chosen from three different locations (similar to the selection used in Fig. \ref{fig:RT_traj}) to study their trajectories. Episodic simulations with long time scales are carried out employing TSAPs for mixing until $\tau = 2\times10^{8}$ time steps. The trajectories of the three representative passive particles are plotted in Fig. \ref{fig:part_pair_evol}(a). These particles are observed to move in roughly circular trajectories around the domain centre, until they reach the area where the TSAPs are active (see Fig. \ref{fig:L_vs_UL}(a-ii)). In the episode presented in Fig. \ref{fig:part_pair_evol}, the passive particles adjacent to the wall move in a counter-clockwise (CCW) fashion, whereas those in the interior regions of the domain perform clockwise (CW) motion. It is also observed that the particles in the path adjacent to the outermost path (band $2$ in Fig. \ref{fig:part_pair_evol}(b)) have a propensity to execute motion in either of the two directions (CW or CCW). Such behaviour can be explained by envisaging the motion of these particles as mimicking the motion of particles kept in between shear layers (moving in opposite directions). To corroborate our findings, basic computational fluid dynamics simulations are carried out to recreate a similar mixing behaviour. The Eulerian simulations utilise two identical fluids differentiated just by colour, with similar initial conditions as the Lagrangian system. Fluid motion (and thereby, mixing) was induced by imposing a constant surface speed on the periphery of an elliptical disk with dimensions close to the operating region of the SAPs. The computational results exhibit analogous mixing dynamics to those observed in the particle system and are further discussed in Sec. SI-4 and Fig. S6 of Supplementary Information.

The behaviour of the passive particles reported in Fig. \ref{fig:part_pair_evol}(a) is representative of the passive particles in all the episodes. This indicates an optimal mixing strategy where the active particles induce a circular motion around the domain centre to promote mixing among initially segregated passive particles. Moreover, the passive particles also have a transverse (radial) component of motion as they move along the circular path. Switching between the concentric paths is predominant in the region where the TSAPs are active (see Fig. \ref{fig:L_vs_UL}(a-ii)) due to frequent collisions with the moving SAPs. To analyse the directionality of the circular motion of the passive particles over a complete trajectory, the domain has been divided into several bands (see Fig. \ref{fig:part_pair_evol}(b); bands are numbered from $1$ to $6$, $1$ being the outermost). Following the trajectory data from numerous passive particles over multiple episodes ($7$ distinct concentric trajectories are observed on superposing all the passive positional data), each band is assumed to have a radial width of $2r$. Figure \ref{fig:part_pair_evol}(b) illustrates the trajectory of a typical particle starting from the periphery (marked by the red disk). The density distribution of the angular velocities within these bands is elucidated in Fig. \ref{fig:part_pair_evol}(c). In the inner bands (bands 3-6), the particle velocities are predominantly in the CW direction (assumed to be negative angular velocity $\omega$; majority of the distribution has $\omega < 0$). In band $2$, the passive particles are almost equally likely to have a CW or CCW bias in their motion, while in band $1$, there is a clear bias towards CCW motion ($\omega >0$). The combined CW and CCW motion in the different regions of the domain culminates in an efficient mixing of the two passive species. 

\subsection*{Effect of $\Gamma$ and $\delta$ on mixing performance}

\begin{figure}[ht]
\centering
\includegraphics[scale=0.85]{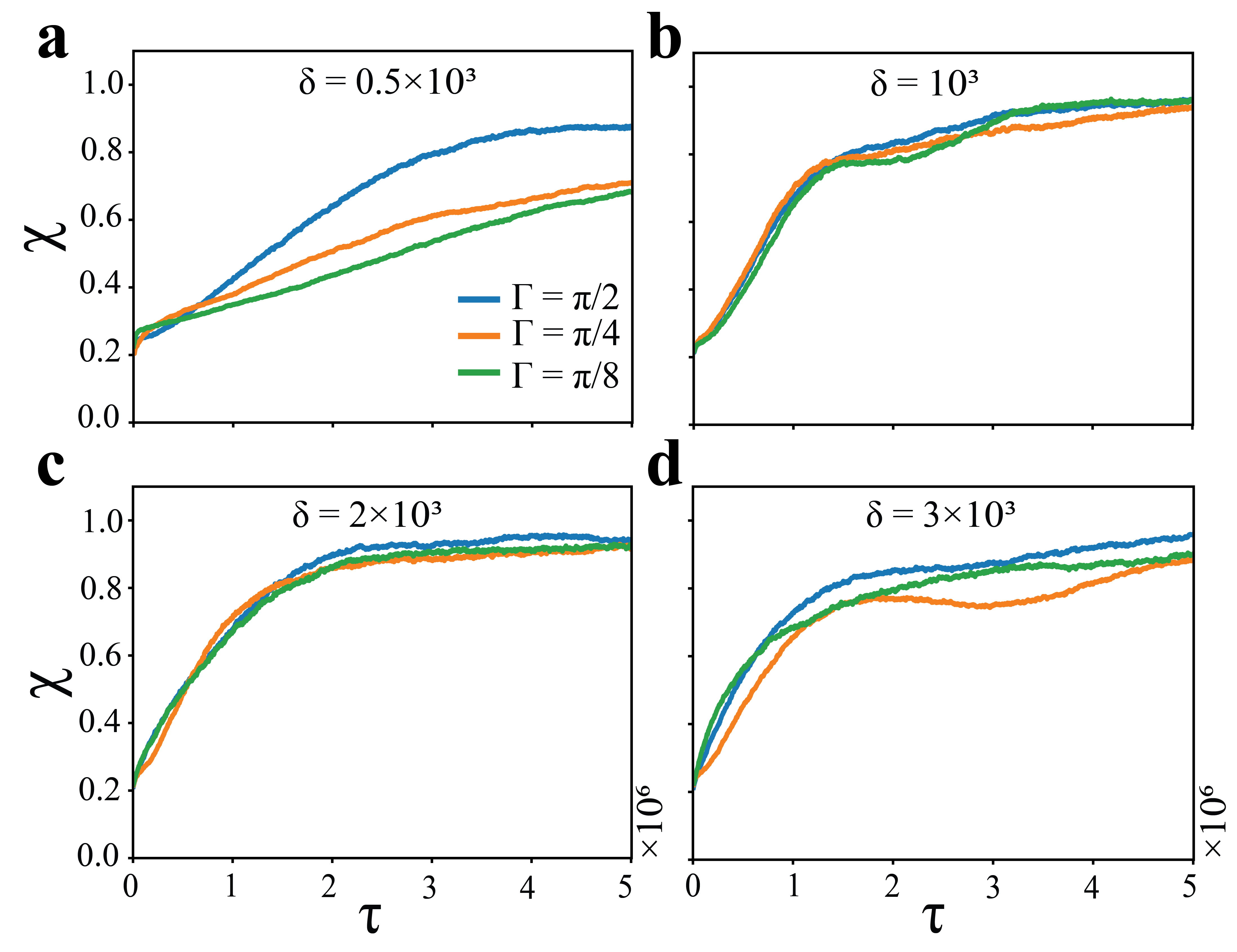}
\caption{The effect of the tumble step $\Gamma$ and run duration $\delta$ of the SAPs is demonstrated with respect to mixing actuated by a trained RL agent. The temporal variation in the mixing index $\chi$ is illustrated at different parametric combinations of $\delta$ and $\Gamma$. Panels (a) through (d) present the mixing index variation for different run durations $\delta=0.5\times10^{3}$ time steps, $\delta=10^{3}$ time steps, $\delta=2\times10^{3}$ time steps, and $3\times10^{3}$ time steps, respectively. In each panel, $\Gamma$ is varied in steps of $\pi/2$, $\pi/4$, and $\pi/8$ (blue, orange, and green curves, respectively). (Note: Each curve is averaged over fifty test episodes. The learning rate for the training of the RL agent is set to $10^{-6}$, and other hyperparameters for the RL algorithm are set to default values defined in the \textit{stable-baselines3} library (refer to Table S1 of Supplementary Information).}
\label{fig:Gamma_effect}
\end{figure}

The run duration $\delta$ and the angle of the tumble $\theta = k\Gamma$ ($k=0,1,2\cdots$) are deemed to be important input parameters governing the active particle dynamics and closely associated with the policy updates during the training of the RL agent. Figure \ref{fig:Gamma_effect} outlines the effect of $\delta$ and $\Gamma$ in the temporal variation of the mixing index $\chi$, when the binary passive system is mixed using trained SAPs. The training of the RL agent for all the combinations of $\delta$ and $\Gamma$ is carried out using a neural network (NN) architecture of hidden layer size ($512,256,64$) and a learning rate of $10^{-6}$. Other hyperparameters used in the PPO algorithm are set to their default values as defined in the \textit{stable-baselines3} package (see Table S1 of Supplementary Information). It is evident from any of the panels (each panel represents a different $\delta$) that $\Gamma=\pi/2$, despite allowing for minimal directional options for the movement of the active particles, is sufficient to induce mixing. Considering a lower $\Gamma$, such as $\Gamma=\pi/4$ or $\Gamma=\pi/8$, increases the angular action space for each active particle, thereby increasing the complexity of the training stage of the RL agent. Meanwhile, the corresponding improvement in the mixing index is nominal in most cases, with adverse effects being observed in certain cases (see Fig. \ref{fig:Gamma_effect}(a)). 

On the other hand, the $\delta$ values used in the training of the RL agent are selected on the basis of the simulations involving the RT particles (see inset of Fig. \ref{fig:op_vs_nop}(c)). It is apparent from Fig. \ref{fig:Gamma_effect}(a) that too low a run duration can lead to sub-optimal mixing even using trained SAPs, if finer $\Gamma$ values are chosen. However, with $\Gamma=\pi/2$, the mixing index $\chi$ saturates to similar peak values ($\chi\approx0.96$), irrespective of the run duration of the SAPs, except at $\delta=0.50\times10^{3}$ for which $\chi\approx0.89$ (value at $\tau=5\times 10^6$). At $\delta=10^{3}$ and $\delta=2\times10^{3}$, the influence of $\Gamma$ is trivial, as all the curves saturate to a similar value following a similar trend (see Fig. \ref{fig:Gamma_effect}(b--c)). However, the variation in the mixing index is smoother in the latter one, and a higher mixing index is obtained at an early stage (hence, a marginal improvement in mixing). Further increase in $\delta$ is found to be detrimental to the mixing performance of the SAPs.

\section*{Discussion} 
\label{sec:Disc}

The current study demonstrates the use of Reinforcement Learning (RL) to train and manage a collection of smart active particles (SAPs) in a high-dimensional state-action space to achieve an optimal mixing between two initially segregated passive species. The forces exerted by the active particles on collision drive the passive particles. The mixing among the two passive species is quantified through a mixing index $\chi$, which is a function of the number fraction of passive particles of opposite species, calculated locally. Extensive simulations show that a discrete action space with SAP movement limited to only four directions is found to suffice for mixing the passive species efficiently. Even with the intrinsic nonlinearity from the inter-particle collisions, the MLP policy employed to capture the best state-action pairs, along with the PPO algorithm (which optimises the RL agent parameters), effectively mimics the coordination among the SAPs. The current work primarily highlights the motion of the SAPs leading to an optimally mixed passive mixture. The operating area of the SAPs in such cases is observed to be fairly restricted to a small area offset from the domain centre, rather than dispersing across the domain, promoting a circular motion among the passive particles about the domain centre. An analogous Eulerian model involving an elliptical-shaped mixer (ellipse-shaped void with constant surface speed) positioned eccentrically in the domain yields a similar area fraction distribution with two immiscible fluids as that observed among the passive particles (refer to Sec. SI-4 and Fig. S6 of Supplementary Information). To demonstrate the efficacy of the active particles controlled by a trained agent, the mixing induced by a set of run-and-tumble (RT) particles with similar tumble angles and run durations has been analysed as a baseline study. From the analysis, the peak $\chi$ increased from $0.90$ for an RT-based system to $0.96$ for an RL-based system with a distinctly faster increase in the latter, clearly emphasising the superiority of the SAPs in mixing the passive system. It is also noted that the dynamics of the RT particles provide findings identical to those generated by NTSAPs (see Sec. SI-5 and Fig. S7 of Supplementary Information). Apart from the mixing index, the mixing performance of the active particles has also been quantified through the probability distribution $p(\zeta)$, $\zeta$ being the ratio of the number of particles of opposite species to the total number of particles around each passive particle in a specified radius $r_{c}$. Using trained SAPs to mix the system yields a Gaussian probability distribution with the peak value close to $\zeta=0.5$, which corresponds to equal numbers of particles of opposite and similar types. At the same time, the distribution in the case of non-trained SAPs exhibits a flatter peak with more negative excess kurtosis, exhibiting an inability to optimise the mixing in the system. To generalise the findings of the work, training and testing of multiple systems with different initial positional distributions for the active and passive particles have been conducted. The obtained results strongly support the applicability of the same RL framework, irrespective of initial particle positions (refer to Sec. SI-6 and Fig. S8 of Supplementary Information).

As the reward system defined in the current work focuses on the maximisation of the mixing index, the RL agent finds an optimum at $\zeta=0.5$. However, a more intricate reward system with additional or different goals can also be tried to assess the effectiveness of integration. A multi-objective reward with suitable weights that balances complementary objectives, such as spatial dispersion, rate of mixing, and penalties for same-species clustering, is worth examining as a future scope. The SAPs used in the current work have constant self-propulsion speeds, which can be added as another parameter to be controlled through continuous or discrete inputs from the agent. Moreover, all the particles interact through an inter-particle collision drive. The current approach of using SAPs can be integrated with existing macroscale techniques involving electric or magnetic fields \cite{diwakar2022ac, shields2017evolution, harraq2022field}. In such cases, attractive and repulsive forces can also be incorporated in the particle dynamics, and the strength of the field can be controlled in tandem with the motion of the SAPs to hasten the mixing process. Such a setup can also find use in segregating a mixed system. 

The system described in the current work can be experimentally realised by substituting the SAPs with either light-activated Janus particles (externally controlled) or micro-robots (either with on-board or off-board actuation mechanisms), with comparable motion characteristics. Fluorescent labelling can be used for real-time tagging/tracking of the motion of the passive species and to distinguish between two or more species. An advantage of the current RL framework is the ease of adapting it to the intended active-passive experimental realisations owing to its modular design. However, experimental realisations of such RL frameworks often struggle with issues related to real-time processing of large volumes of data, which can result in delayed feedback and inaccurate state inferences. Additionally, data might be noisy or partial, which makes it challenging for RL algorithms to properly figure out the real state of the environment that is needed to make the best decisions. Leveraging a trained RL model from a minimalist model, as implemented here, can serve as a robust initialiser, significantly accelerating optimisation in the real environment compared to training a policy entirely from scratch.

Furthermore, the findings reported in the current work have significant ramifications for the study of controllable active matter systems. Owing to the generality of the micro-robotic system, the same system can be put to use in various fields simply by redefining the agent's input parameters, reward function, and governing equations while choosing the appropriate neural networks and hyperparameters. By adjusting the objective function, these smart active particles can be used in a variety of fields, such as targeted drug delivery \cite{park2017multifunctional}, microswimmer-based mixing \cite{bailey2024low}, smart navigation in colloidal and complex environments \cite{yang2020efficient}, and granular mixing and segregation \cite{agrawal2021alignment}, where active particles interact and regulate passive entities. To enhance the realism of the current work, a logical next step could be the incorporation of polydispersity in the particle properties. In this scenario, the RL agent must be aware of both the particle positions and their sizes, along with their identities. Such modification, accompanied by the refinement of the reward mechanism and the agent architecture, can further the functionality and practicality of the SAPs. Augmenting the system to three dimensions could enhance the versatility; however, at the cost of substantially increasing the mobility, interactions, and trajectories, necessitating three-dimensional spatial information for training the agent, which translates to much higher computational costs. Finally, by offering an adaptable framework that can be adjusted for active particles involved in multi-body interactions, the current work promotes the integration of conventional active matter theory with powerful reinforcement learning techniques.
 
\bibliography{References_v2}

\section*{Acknowledgments}

This research is partially supported by the Indian Institute of Technology Madras [Sanction No. SB22231233MEETWO008509, RF22230093MERFIR008846]. PSM acknowledges the V. Ganesan Faculty Fellowship received from IIT Madras.

\section*{Author contributions statement}
TJ, SiM and PSM contributed to conceiving and developing the idea. TJ, SiM and SaM developed the code. TJ ran the simulations and wrote the initial draft of the manuscript. RA developed the analogical model. TJ and SiM post-processed the data. All authors reviewed the manuscript. PSM funded the work.

\section*{Additional information}
The simulation code and the data for making the plots can be found at \url{https://github.com/s-m-sys/mix_with_SAPs}.

\end{document}